\documentclass{aa}  
\usepackage{multirow}
\usepackage{graphicx}
\usepackage{txfonts}
\usepackage{lipsum}
\usepackage{subcaption}         
\usepackage{lscape}             
\usepackage{placeins}           
\usepackage{comment}
\usepackage{xcolor}
\usepackage[normalem]{ulem} 

\usepackage[]{hyperref}

\newcommand{\Ms}{ \mathrm{M}_\odot }

\begin{document}

\title{Expanding stellar associations as Galactic accelerometers}

\author{Till Sawala\inst{1}\corrauth{till.sawala@helsinki.fi}  
\and Núria Miret Roig\inst{2,3}\email{ nuria.miret.roig@icc.ub.edu} }
\institute{Department of Physics, University of Helsinki, Gustaf H\"allstr\"omin katu 2, FI-00014 Helsinki, Finland
\and Dep. de Física Quàntica i Astrofísica (FQA), Univ. de Barcelona (UB), Martí i Franquès, 1, 08028 Barcelona, Spain
\and Institut de Ciències del Cosmos (ICCUB), Univ. de Barcelona (UB), Martí i Franquès, 1, 08028 Barcelona, Spain}

\date{ }
 
\abstract
{The gravitational potential of the Milky Way is fundamental for understanding the evolution of our Galaxy and the nature of dark matter.}
{We introduce a new method to constrain the Galactic potential using expanding young stellar associations. We exploit the physical constraint that these stars share a common, compact origin to reconstruct their most likely orbits and infer the gravitational potential in which they have evolved.}
{We define the size of an association using the trace and determinant of its position covariance matrix. By integrating synthetic associations backward in different trial potentials, we show how the true potential can be identified as the one that minimises these metrics.}
{We demonstrate that, while current observational errors are still too large, upcoming observations will allow us to distinguish between different potentials. Our results suggest that with Gaia DR4 astrometry and radial velocity errors below $0.2\,\mathrm{km\,s^{-1}}$, the halo mass can be constrained with a precision of $< 0.6 \times 10^{12}~\Ms$ and the concentration with a precision of $< 0.8$ using a single association, albeit with significant degeneracies. In addition, we show that the inferred dynamical traceback age is sensitive to the gravitational potential, suggesting that independent age information can help break existing degeneracies, but also that traceback-age estimates are not independent of the assumed potential.}
{Expanding stellar associations carry information about the gravitational potential in which they have evolved. With the arrival of next-generation astrometry and high-precision radial velocities, they will provide a complementary tool for constraining the Galactic potential.}


\maketitle
\nolinenumbers

\section{Introduction}
The total mass of the Milky Way and the shape of its gravitational potential are fundamental variables in galactic dynamics and cosmology, linked to questions such as the nature of dark matter \citep[e.g.][]{Kennedy-2014, Sawala-2016a, Leane-2020}, the physics of adiabatic contraction \citep[e.g.][]{Hussein-2025}, and even the future fate of the Local Group \citep{Sawala-2025}.
Due to this central role and the abundance of observational data, the Galactic potential has been probed using many different techniques and tracers, including disk stars \citep[e.g.][]{Cautun-2020, Karukes-2020, Ablimit-2020, Shen-2022} and halo stars \citep[e.g.][]{Watkins-2019}, globular clusters \citep[e.g.][]{Watkins-2019, Posti-2019}, high velocity RR-Lyrae stars \citep{Prudil-2022}, and satellite galaxies \citep[e.g.][]{Callingham-2019, Fritz-2020, Li-2020, Rodriguez-Wimberly-2022}. Recent overviews are given, for example, by \cite{Wang-2020, Sawala-2023a, Hayati-2024}, and \cite{Hunt-2025}.

Because tracer particles are generally well mixed, inferring the gravitational acceleration from present-day phase–space coordinates alone requires additional assumptions (e.g. on orbital anisotropy or distribution functions), which introduce degeneracies with the underlying potential. Some tracers, however, possess additional constraints that can help anchor their orbits and break degeneracies. One example is stars belonging to stellar streams \citep[e.g.][]{Johnston-1999, Bovy-2016, Ibata-2017, Mateu-2023, Nibauer-2025, Bonaca-2025}, which have been deposited along the orbit of a tidally disrupted system, such as a globular cluster or dwarf galaxy. The dependence of the deposition of stars on the potential, and the fact that stream members originate from a small region in phase-space, can be used to `rewind' their trajectories, revealing the potential in which the stream has formed and evolved \citep[e.g.][]{PW-2014, Gibbons-2014, Palau-2025}. In practice, this is done by forward-modelling stream formation in trial potentials and fitting to the stream’s track, width and kinematics.

In this work, we propose to use stars belonging to individual expanding young associations as tracer particles. In this case, the fact that all stars belonging to the same association have originated from a compact birth region \citep[e.g.][]{Ambartsumian-1947,Gieles-2011,CantatGaudin-2019} provides the missing constraint that allows a full orbital reconstruction, which can then be used to infer the underlying potential.
The method of tracing back expanding stellar associations to their birth has previously been exploited successfully to determine the dynamical traceback age of the stellar association \cite[e.g.][]{Song-2003, Fernandez-2008, Miret-Roig-2018, Miret-Roig-2020, Miret-Roig+2022, Kerr+2022, Couture+2023, Couture-2026, Galli-2023, Galli-2024}. We explore the ability to determine accurate dynamical traceback ages based on current and future observational capabilities, and introduce a new method to constrain  the Galactic potential.

\begin{figure*}
\centering
   \includegraphics[width=7in]{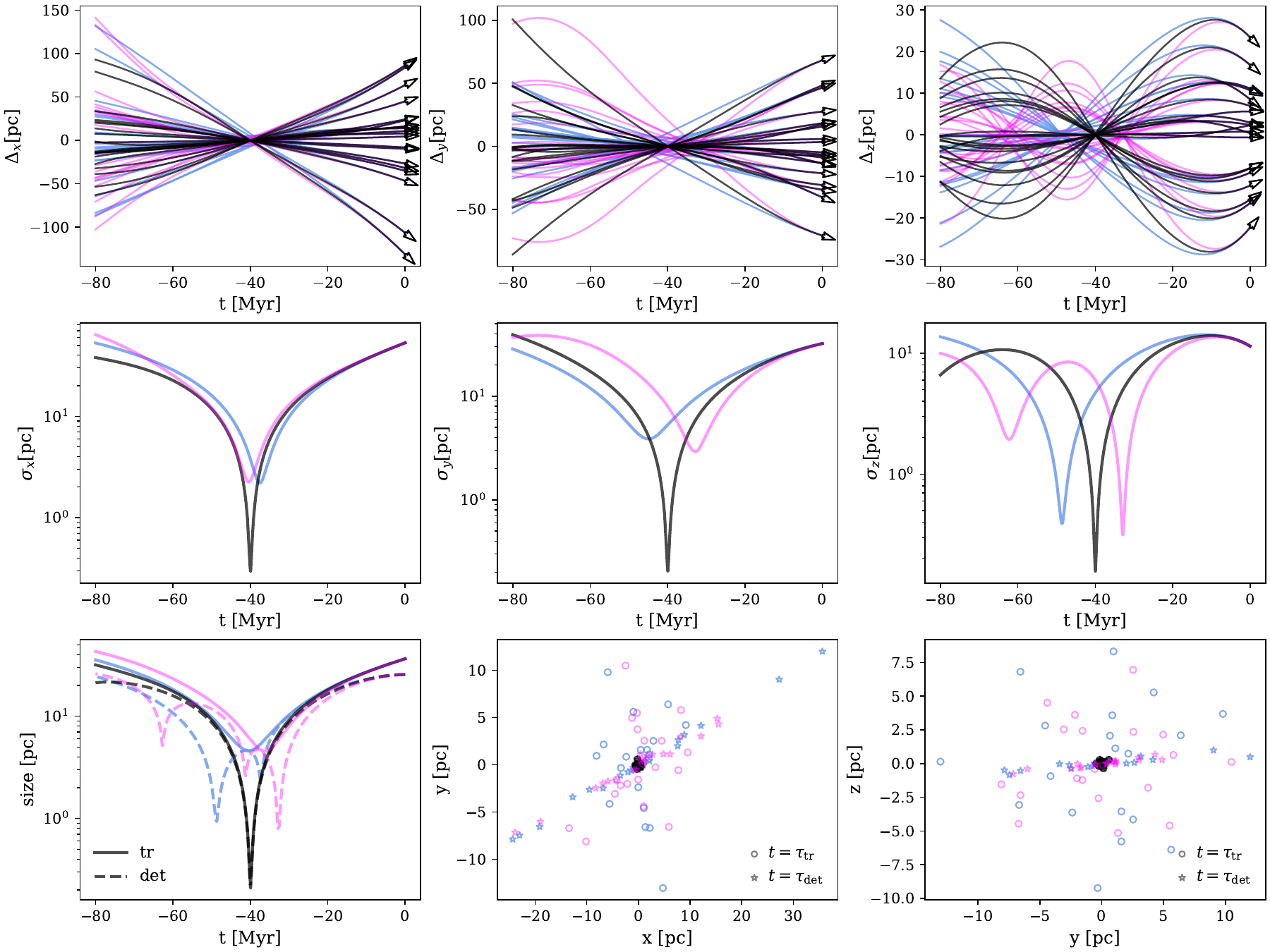}
    \caption{Illustration of the method. An association of $N_\star=20$ stars is integrated backward from identical phase–space coordinates at $t=0$ in three Galactic potentials: the true potential (black) and two alternatives (red, blue). The top row shows stellar positions relative to the measured centre in Galactic $x$, $y$, and $z$ coordinates; the middle row shows the corresponding dispersions. The bottom-left panel shows the evolution of the trace, $\mathrm{tr}(t)$, and the determinant, $\mathrm{det}(t)$, of the position covariance; the bottom-middle and bottom-right panels show the stellar positions at $t = \tau_\mathrm{tr}$ and $t = \tau_\mathrm{det}$, i.e. the times when $\mathrm{tr}(t)$ and $\mathrm{det}(t)$ are minimal. In the true potential, both metrics reach the smallest minimum at the association's formation time. By evolving an observed association in different potentials and comparing the minima of $\mathrm{tr}(t)$ or $\mathrm{det}(t)$, the true potential can be inferred.}
    \label{fig:illustration-method}
\end{figure*}

\section{Methods}

\subsection{Definitions}~\label{sec:definition}
Throughout this work, we parametrise the size of the association using either the trace or the determinant of the covariance matrix, $\bf{C}$, of the Cartesian positions of the individual stars. In each dimension, we define the variance, $V(x)=\frac{\sum_{i=1}^{N_\star}(x_i-\mu_x)^2}{N_\star}$, and for any pair of dimensions, the covariance, $Cov(x,y)=\frac{\sum_{i=1}^{N_\star}[(x_i-\mu_x)(y_i-\mu_y)]}{{N_\star}}$, with $Cov(x,y) = Cov(y,x)$ and $Cov(x,x) = V(x)$. The covariance matrix is then 

$$
\bf{C}=\begin{bmatrix}
V(x) & Cov(x,y) & Cov(x,z) \\ \\
Cov(x,y) & V(y) & Cov(y,z) \\ \\
Cov(x,z) & Cov(y,z) & V(z) \\ \\
\end{bmatrix}.
$$
The trace and determinant of $\bf{C}$ are
$$\mathrm{Tr} = V(x) + V(y) + V(z),$$ 
and
$$
  \begin{aligned}
    \mathrm{Det} & = V(x) \cdot V(y) \cdot V(z)\\
      & - V(x) \cdot Cov(y,z)^2 \\
      & - V(y) \cdot Cov(x,z)^2 \\
      & - V(z) \cdot Cov(x,y)^2 \\
      & + 2 \ Cov(x,y) \cdot Cov(x,z) \cdot Cov (y,z),\\
  \end{aligned}
$$
respectively. Because the off-diagonal terms of $\bf{C}$ are typically small and enter only at second or higher order, \mbox{$\mathrm{Det} \sim V(x) \cdot V(y) \cdot V(z)$}. In other words, $\mathrm{Tr}$ is the sum of the variances, while 
$\mathrm{Det}$ approximates their product in three dimensions.

In the following, we use $\mathrm{tr} \equiv \mathrm{Tr}^{1/2}$ and $\mathrm{det}  \equiv \mathrm{Det}^{1/6}$ for the associated linear quantities, which capture complementary notions of an association's size.
The trace is sensitive to any single direction becoming large, and tends to behave smoothly under measurement noise because it adds contributions linearly. The determinant is sensitive to contraction in any dimension, and can become small even when the configuration is pancake-like or cigar-like, if at least one principal axis is small.  It can be a sharper discriminator of true 3D focusing, but it is also more numerically stiff and more sensitive to sample variance and outliers.

We define the dynamical traceback ages, $\tau_\mathrm{tr}$ and $\tau_\mathrm{det}$, as the times at which either $\mathrm{tr}(t)$ or $\mathrm{det}(t)$ attain their minima, i.e.
$$
\begin{aligned}
\tau_\mathrm{tr} &\equiv \operatorname*{arg\,min}\!\left[\mathrm{tr}(t)\right], \\
\tau_\mathrm{det} &\equiv \operatorname*{arg\,min}\!\left[\mathrm{det}(t)\right],
\end{aligned}
$$
and we use the values of the minima, 
$$
\begin{aligned}
\mathrm{tr}_\mathrm{min} &\equiv \min\!\left[\mathrm{tr}(t)\right],\\
\mathrm{det}_\mathrm{min} &\equiv \min\!\left[\mathrm{det}(t)\right],
\end{aligned}
$$
as objective functions to infer the potential. We discuss their differing behaviour in practical examples in the following sections.

\subsection{Concept}
The main idea behind our method is illustrated in Figure~\ref{fig:illustration-method}, which shows the evolution of a synthetic association comprising $N_\star = 20$ stars, using the same set of observed phase-space coordinates in each case but integrating the stars backwards in three different Galactic potentials. If, as we assume here, the stars were born in a region much smaller than their present extent, they reach a similarly compact size only when integrated backwards in the true potential, i.e. in the same potential in which they have evolved from their birth to the present time.

The expansion of a stellar association is determined by its initial spatial extent and velocity dispersion, its spatial extent and velocity distribution over time, and the local tidal tensor, which determines the differential acceleration. This tidal tensor, which depends on the gravitational potential, also changes along the orbital path of the association, which is itself determined both by the initial position and velocity of the association, and the gravitational potential. In order to infer the potential, we rely on the dependence of both the orbit and the tidal tensor on the underlying potential, using a parametric density profile.

\begin{figure*}
\centering
   \includegraphics[clip,trim=3cm 0.5cm 3cm 1.9cm, width=18.3cm]{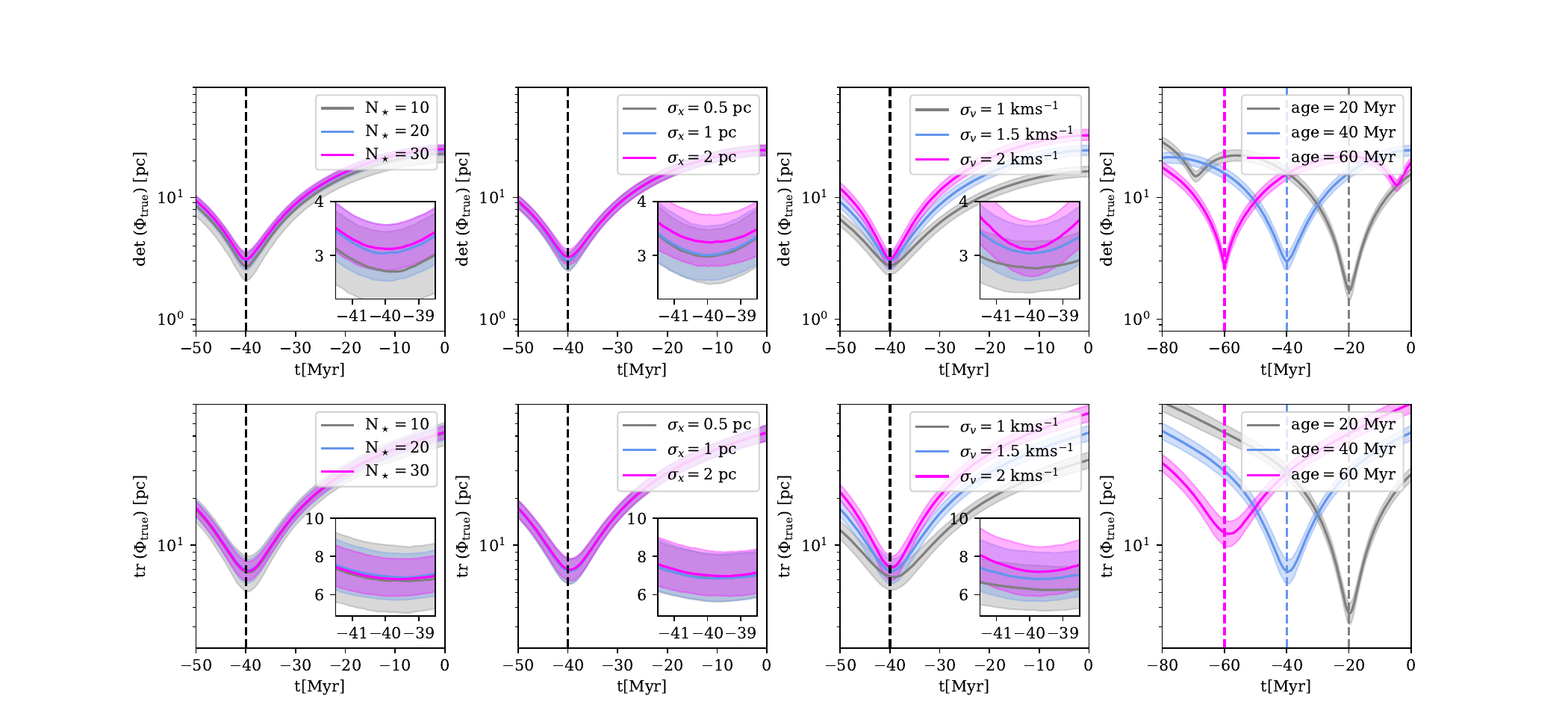}
    \caption{Effect of intrinsic properties on an association's reconstructed evolution in the true potential. The top row uses the determinant of the covariance matrix. The bottom row uses its trace. From left to right, we vary the number of stars, $N_\star$, the initial dispersions in position, $\sigma_x$, and in velocity, $\sigma_v$, and the age of the association. All other properties are set to the default values. Solid lines denote medians, bands indicate $\pm 1 \sigma$ uncertainties obtained by randomly sampling the `next' observational errors and the intrinsic dispersions in positions and velocities. Dashed vertical lines indicate the true age of the association. Insets use linear axes.}
    \label{fig:time-intrinsic}
\end{figure*}

\begin{figure*}
\centering
   \includegraphics[clip,trim=3cm 0.5cm 3cm 1.7cm,width=18.2cm]{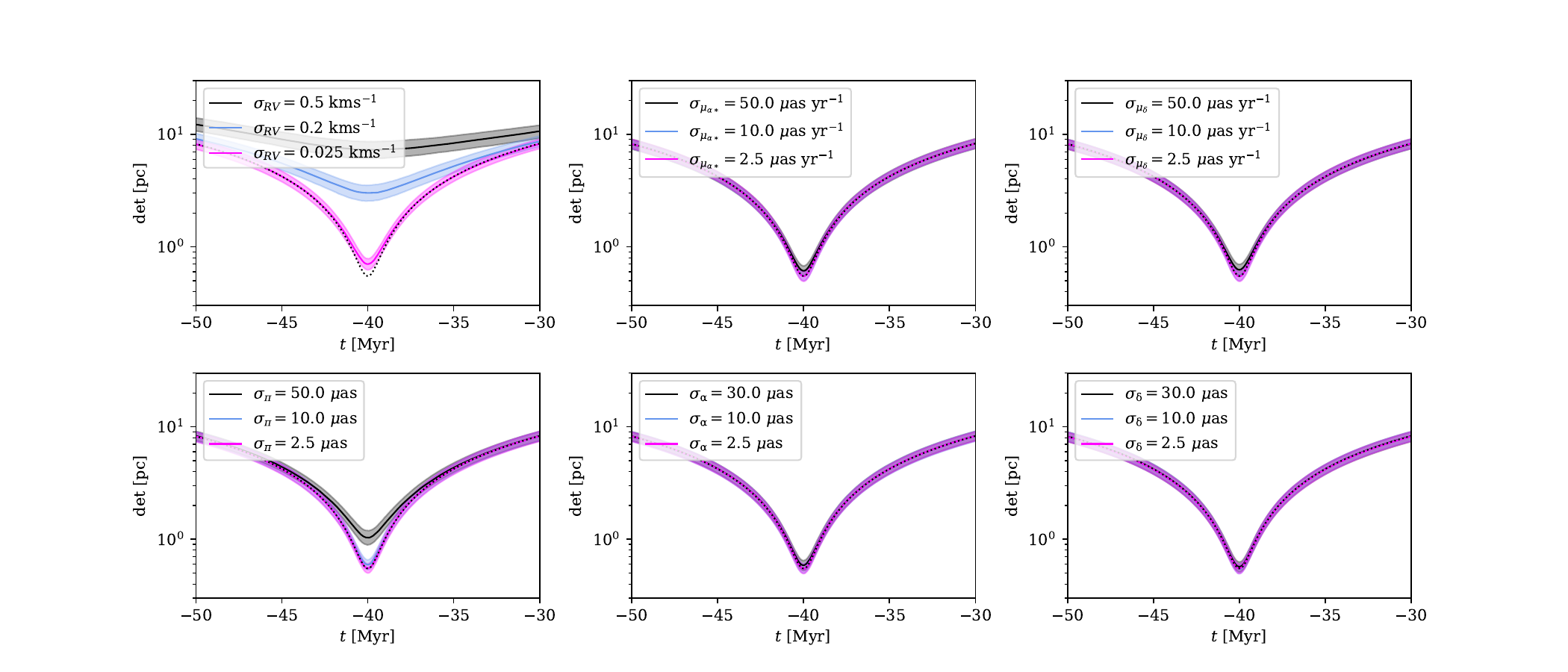}
   \includegraphics[clip,trim=3cm 0.5cm 3cm 1.7cm,width=18.2cm]{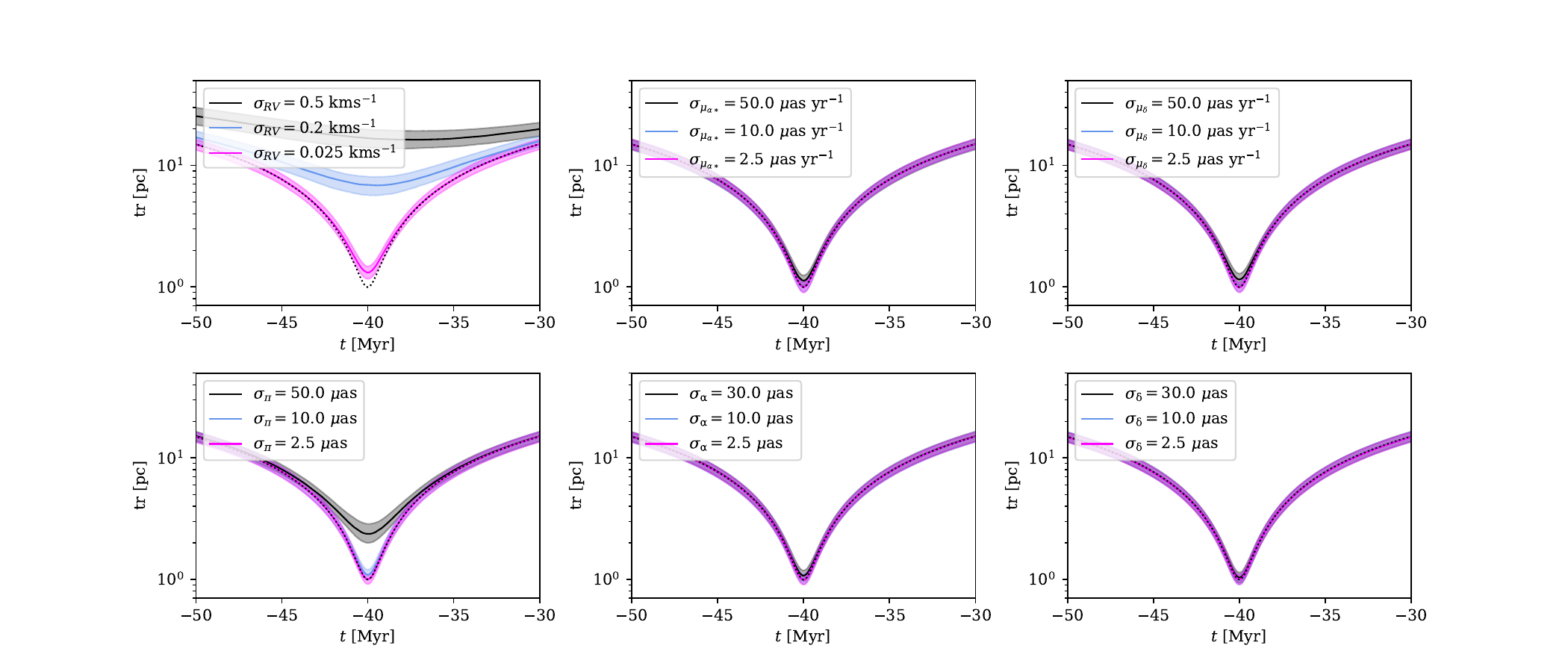}
    \caption{Effect of individual observational errors on an association's reconstructed evolution in the true potential. The top two rows use the determinant of the covariance matrix, the bottom two rows use its trace. Solid lines denote medians, bands indicate $\pm 1 \sigma$ uncertainties. Three error scenarios are considered, namely, `current' (black), `next' (blue),  and `future' (pink), with all other errors set to zero. We also show, as dotted lines, the averages of realisations with zero measurement errors.}
    \label{fig:time-errors}
\end{figure*}

Starting from identical coordinates and velocities at $t=0$, the sets of trajectories of member stars integrated in different potentials begin to diverge with lookback time. While, in this case, a minimum size is reached in all three cases, the values of the minima, $\mathrm{tr}_\mathrm{min}$ and $\mathrm{det}_\mathrm{min}$ and the times at which these minima are reached (i.e. the dynamical traceback ages, $\tau_\mathrm{tr}$ and $\tau_\mathrm{det}$, respectively), depend on the assumed potential. We can identify the true potential and the corresponding true traceback age as those with the smallest value of $\mathrm{tr}_\mathrm{min}$ or $\mathrm{det}_\mathrm{min}$.

It is worth noting again that the same observed positions and velocities of all stars at $t=0$ are used irrespective of the assumed potential. The mere observation of their phase-space coordinates would not provide sufficient information to determine the potential in which they evolved. It is only thanks to the existence of a common origin that the potential can be determined.

We also note that the minimum size generally occurs at different times in different potentials. We discuss below to what extent the traceback age is sensitive to the assumed potential. Furthermore, we note that in order to determine the potential, we do not need prior knowledge of the time at which the association was born, as the dynamical traceback age is simultaneously determined. However, as we discuss later, independent knowledge of an association's age could also be used as additional constraining information to determine the potential.

Finally, we note that, in the two alternative potentials, the time of the minimum trace, $\tau_\mathrm{tr}$, and the time of the minimum determinant, $\tau_\mathrm{det}$, are not identical. It can be seen that, in both alternative potentials, at $\tau_\mathrm{tr}$, the shape of the association is fairly isotropic, while at $\tau_\mathrm{det}$, it is highly anisotropic, cigar-shaped. This can be understood considering the different definitions of $\mathrm{tr}$ and $\mathrm{det}$ described in Section~\ref{sec:definition}: at $\tau_\mathrm{tr}$, the sum of the variances of principal coordinates is minimal, while at $\tau_\mathrm{det}$, their product is minimal.

\section{Inference from synthetic associations}~\label{sec:synthetic}
To demonstrate and test the method under different, more general assumptions, we generated a set of synthetic associations, for which we specified the number of stars, $\mathrm{N}_\star$, the initial size, $\sigma_x$, velocity dispersion, $\sigma_v$, and the age (the effects of these are discussed in Section~\ref{sec:intrinsic}). We also specified the present position and velocity of the association's centre within the Galaxy and relative to the Sun in terms of the following observables: right ascension, $\alpha$, declination, $\delta$, parallax, $\pi$, radial (line of sight) velocity, $RV$, and proper motions, in right ascension multiplied by $\mathrm{cos \ \delta}$, $\mu_{\alpha \ast} \equiv \mu_{\alpha} \ \mathrm{cos \ \delta}$, and in declination, $\mu_\delta$. In our fiducial model, we used the present-day position, radial velocity, and proper motions of the centre of the Tucana-Horologium association (hereafter THA): $\alpha=40.45^\circ$,
$\delta= -53.00^\circ$,
$\pi= 22.65$~mas, $RV = 6.89\ \mathrm{km \ s}^{-1}$,
$\mu_{\alpha \ast} = 88.55\ \mathrm{mas \ yr}^{-1}$, $\mu_\delta = -56.16 \ \mathrm{mas \ yr}^{-1}$ \citep{Galli-2023}. The procedure can be easily applied to other associations in the solar neighbourhood. We repeated the procedure using the mean coordinates and velocities of the $\beta$-Pictoris association (hereafter $\beta$-Pic): $\alpha= 273.21^\circ$, $\delta= -52.08^\circ$, $\pi= 20.34$~mas, $RV = 0.27 \mathrm{\ km \ s}^{-1}$, $\mu_{\alpha \ast} = 13.19\ \mathrm{mas \ yr}^{-1}$, $\mu_\delta = -70.21\  \mathrm{mas \ yr}^{-1}$ \citep{Miret-Roig-2020}. The results are very similar to those obtained for THA. 

Having defined these properties, we integrated the observed position and velocity of the association's centre backwards to a time corresponding to its specified age in the assumed true potential. We then generated a set of individual positions and velocities for all $\mathrm{N}_\star$ stars, with specified dispersions $\sigma_x$, and $\sigma_v$, respectively, centred on the position and velocity of the association's centre at birth. Next, we integrated the orbits of all stars forward to the present time in the same true potential, where we finally convolved all observables with the assumed observational errors (see Section~\ref{sec:errors}). The resulting positions and velocities at $t=0$ constituted our mock observables.
To demonstrate the ability to infer the gravitational potential, we then performed the backwards integration again in different potentials, as described in Section~\ref{sec:potential}. 

Throughout this work, we assumed that the potential was static and consisted of an NFW dark matter halo \citep{NFW-1996}, specified by its mass, $\mathrm{M_h}$, and concentration, $c_\mathrm{NFW}$, a Miyamoto-Nagai stellar disk \citep{Miyamoto-1975} of mass $\mathrm{M_d}$, a bulge of mass $\mathrm{M_b}$, and a nucleus of mass $\mathrm{M_n}$. All orbital integrations were performed numerically using \texttt{Gala} \citep{Price-Whelan-2017}.

In the absence of measurement errors, as illustrated in Figure~\ref{fig:illustration-method}, the phase-space coordinates of a handful of stars are sufficient to reconstruct their orbits based on the knowledge of a common origin, and to uniquely determine both the true potential and the true age. In practice, the ability to reconstruct the orbits and infer the potential depends on both the intrinsic properties of the association, discussed in Section~\ref{sec:intrinsic}, and on the measurement errors, discussed in Section~\ref{sec:errors}.

In the idealised example shown in Figure~\ref{fig:illustration-method}, both the minimum trace and the minimum determinant identify the correct potential and the correct traceback age. However, already in this example, the sensitivity to the potential differs between the two metrics. In the following sections, we present results in terms of both $\mathrm{tr}$ and $\mathrm{det}$, and we compare their sensitivity in Section~\ref{sec:potential}.

\subsection{Effects of the intrinsic properties of the association}\label{sec:intrinsic}
Expanding stellar associations possess several intrinsic properties that affect orbital reconstruction. In this section, we explore the effects of the number of stars, $\mathrm{N}_\star$, size at birth, $\sigma_x$, velocity dispersion at birth, $\sigma_v$, and the true age of the association. Table~\ref{tab:intrinsic} lists the default values of all parameters, as well as two alternative values explored.

We considered associations with very few members because we included only members with complete and precise 6D phase-space information. Many associations have several tens or hundreds of known members with precise 5D astrometry. However, to date, only a few have radial velocities precise enough for an accurate orbit reconstruction. The size and velocity dispersion at birth are similar to those of young associations in the solar neighbourhood. Since orbital errors increase with time integration, we focused on associations younger than 60~Myr, where precision is highest.

\begin{table}
\caption{Intrinsic properties of synthetic associations}
\vspace{-3mm}
\label{tab:intrinsic}
\begin{center}
\renewcommand{\arraystretch}{1.4}
\setlength{\tabcolsep}{15pt}
\begin{tabular}{|l| c c c |} 
\hline 
&  low& default   & high\\  \hline
$\mathrm{N_\star}$ & 10 & 20 & 30 \\ 
$\sigma_x$ &  0.5 pc & 1 pc  & 2  pc\\ 
$\sigma_v$ & $1 \ \mathrm{km \ s}^{-1}$ & $1.5 \ \mathrm{km \ s}^{-1}$ & $2 \ \mathrm{km \ s}^{-1}$\\  [0.5ex]
Age & 20 Myr & 40 Myr & 60 Myr \\ 
 \hline
\end{tabular}
\end{center}
\tablefoot{
The central column lists the values of the default association, which are used unless otherwise stated. The left and right columns show alternative values explored in Section~\ref{sec:intrinsic}.
}

\end{table}

Throughout this section, we adopted the observational errors expected for Gaia DR4, complemented by precise radial velocities from ground-based spectroscopic surveys. These errors are labelled as `next' in Table~\ref{tab:errors}. To sample the observational errors, we created Monte Carlo (MC) samples and integrated each association both forwards and backwards only in the true gravitational potential. 

Figure~\ref{fig:time-intrinsic} shows the size evolution of associations generated with different intrinsic properties. The association with default parameters is represented in blue. The two other choices, labelled as low and high parameters in Table~\ref{tab:intrinsic}, are represented by pink and grey colours. In the top row, we show the time-evolution of the determinant of the covariance, while in the bottom row, we show the evolution of the trace.

The insets show the region around the minima, $\mathrm{det}_\mathrm{min}$ and $\mathrm{tr}_\mathrm{min}$. What matters for inference is not simply the absolute value of the minimum, but how strongly the expected minimum changes with the parameter of interest relative to its scatter across Monte Carlo realisations, which in turn depend on the observational errors. A steeper dependence and a smaller scatter yield tighter constraints on the age or the potential.

As expected, associations with fewer stars have smaller measured sizes on average (see the first column of Figure~\ref{fig:time-intrinsic}). They also have greater variation of the measured size among the MC samples, which is reflected in greater uncertainties on the minimum sizes, $\mathrm{det}_\mathrm{min}$ and $\mathrm{tr}_\mathrm{min}$, and on the ages, $\tau_\mathrm{det}$ and $\tau_\mathrm{tr}$.

A smaller spatial dispersion at birth, $\sigma_x$, leads to smaller minimum sizes (see the second column of Figure~\ref{fig:time-intrinsic}). However, for the range of parameters and errors considered, reducing the assumed spatial dispersion below $\sigma_x = 1$~pc does not strongly impact the inferred minimum size, which is limited by the observational errors.

The intrinsic velocity dispersion, $\sigma_v$, and the true age of the association affect the reconstruction in more important and slightly less obvious ways. Associations born with smaller velocity dispersions can be traced back to slightly smaller minimum sizes on average (see the third column of Figure~\ref{fig:time-intrinsic}). However, because they also grow to smaller sizes at $t=0$, their inferred expansion is more affected by observational errors, and both their minimum sizes and times of the minima are less clearly defined than those of associations with larger intrinsic velocity dispersions. 

As shown in the fourth column of Figure~\ref{fig:time-intrinsic}, younger associations with the same intrinsic properties can also be traced back to smaller sizes than older ones, allowing a more precise absolute age estimate. However, because the stars in older associations have evolved in the external potential for longer, the imprint of varying potentials can be more readily discerned in older associations (see Sect.~\ref{sec:potential}).

\subsection{Sensitivity to observational errors} ~\label{sec:errors}
Observational errors limit the ability to reconstruct individual stellar orbits. For our purpose, measurement errors effectively reduce the convergence of orbits at birth, which limits the ability to measure the age or distinguish the true potential from alternatives.

\begin{table}
\caption{Observational errors}
\vspace{-3mm}
\label{tab:errors}
\begin{center}
\renewcommand{\arraystretch}{1.4}
\setlength{\tabcolsep}{11pt}
\begin{tabular}{|l| r r r|} 
\hline 
&  `current' & `next' & `future'\\  [0.5ex]  \hline
$\sigma_{\alpha}$ & 30 $\mu$as & 10 $\mu$as & 2.5 $\mu$as\\ 
$\sigma_{\delta}$ &  30 $\mu$as & 10 $\mu$as & 2.5 $\mu$as\\ 
$\sigma_{\pi}$ & 50 $\mu$as & 10 $\mu$as & 2.5 $\mu$as\\  [0.5ex]
$\sigma_{\mu_{\alpha \ast}}$ & $50\ \mu \mathrm{as\ yr}^{-1}$ & $10\ \mu \mathrm{as\ yr}^{-1}$  & $2.5 \ \mu \mathrm{as\ yr}^{-1}$\\ 
$\sigma_{\mu_{\delta}}$ & $50 \ \mu \mathrm{as\ yr}^{-1}$ & $10 \ \mu \mathrm{as\ yr}^{-1}$  & $2.5 \ \mu \mathrm{as\ yr}^{-1}$\\ [0.5ex]
$\sigma_{RV}$ & $0.5\ \mathrm{km \ s}^{-1}$ & $0.2 \ \mathrm{km \ s}^{-1}$ & $0.025 \ \mathrm{km \ s}^{-1}$ \\ 
 \hline
\end{tabular}
\end{center}
\tablefoot{
The central column lists `next' errors, which are used unless otherwise stated. The left and right columns show combinations of larger (`current') or smaller (`future') errors, explored in Section~\ref{sec:errors}.
}
\end{table}

We added independent Gaussian uncertainties to all observables $(\alpha, \delta, \pi, \mu_{\alpha*},\mu_\delta, RV)$ of individual stars, using the three error budgets labelled as `current', `next', and `future' and listed in Table~\ref{tab:errors}. `current' observational errors represent the state of the art, based on Gaia DR3 astrometry for parallax, $\pi$, sky coordinates, $\alpha$ and $\delta$, and proper motions, $\mu_{\alpha \ast}$ and ${\mu_{\delta}}$, and on complementary spectroscopic observations of the radial velocity, $RV$. The errors in the `next' scenario assume the improvement in astrometry predicted by Gaia DR4\footnote{\url{https://great.ast.cam.ac.uk/Greatwiki/GreatMeet-PM18?action=AttachFile&do=view&target=EAS2025-S1-Brown.pdf}} and slightly reduced radial-velocity errors. The `future' errors envision further improvements beyond Gaia DR5, such as those expected by early data releases of Gaia-NIR\footnote{\url{https://indico.icc.ub.edu/event/675/contributions/4823/attachments/2032/4021/Seville_2026_GaiaNIR_Hobbs.pdf}} accompanied by a significant improvement in radial-velocity precision. In the following sections, we abbreviate the `current', `next', and `future' errors as $\epsilon_c$, $\epsilon_n$, and $\epsilon_f$, respectively.

In Figure~\ref{fig:time-errors}, we show the effect of the individual observational errors on the reconstruction of associations with the default combination of intrinsic properties (see Table~\ref{tab:intrinsic}). We show the average evolution of either the determinant (top two rows) or trace (bottom two rows) of the covariance matrix for the default association, integrated forwards and backwards in the default true potential. Observational errors were applied at the present time ($t=0$). On each panel, we considered the three error scenarios on a different variable, keeping the errors on the other variables set to zero. 

The dotted lines indicate the average of the inferred evolution assuming no measurement errors (but sampling many realisations of the default association). Coloured lines show the average evolution inferred after applying the corresponding errors to each sample, while bands show percentiles corresponding to $\pm 1 \sigma$ errors.

It can be seen that even very small measurement errors in any of the variables introduce a notable scatter in the determinant. This is caused by the fact that the determinant is set by the product of three eigenvalues, making it numerically stiff and sensitive to the shape of the particular association (as illustrated in Figure~\ref{fig:illustration-method}). This introduces a large degree of sample variance, in addition to and independent of measurement errors. However, as discussed in Section~\ref{sec:age-inference}, this affects the value of $\mathrm{det}$ and of its minimum, $\mathrm{det}_\mathrm{min}$, more strongly than the time $\tau_\mathrm{det}$ at which the determinant is minimal in individual associations, making $\tau_\mathrm{det}$ a common and valid choice for traceback ages. By contrast, the trace shows much less sensitivity and smaller sample variance.

For associations in the solar neighbourhood, the most significant source of errors is the radial-velocity measurement. The impact of a `current' parallax error of 50 $\mu$as is comparable with that of the `next' radial-velocity error. 
To profit from the improved astrometry of future Gaia releases, it is crucial to achieve equivalent improvements in the RV precision. At these distances, the parallax errors are subdominant. The errors in the sky coordinates and proper motions are insignificant even at their `current' values.

However, at larger distances, the effect of the parallax error increases significantly. Beyond 200~pc ($\pi<5$~mas), the parallax error begins to dominate, and proper motion errors start to become important. This limits the analysis to the solar neighbourhood in the present and immediate future, but could be extended to other regions, based on significant improvements in astrometry expected by future missions like GaiaNIR \citep{Hobbs-2016, Hobbs-2021}.

Comparing the evolution of the determinant and the trace in the top and bottom two rows of Figure~\ref{fig:time-errors}, respectively, reveals clear differences: while for the trace, the spread between individual samples of associations reduces along with the errors, for the determinant, there is a notable sample variance even as the observational errors tend to zero. This residual sample variance is caused by the finite variance in the determinant of the covariance matrix of isotropic random realisations of associations with moderate numbers of stars (in this example, $N=20$).

\begin{table}
\caption{Parameters of the Galactic potential}
\vspace{-3mm}
\label{tab:potential}
\begin{center}
\renewcommand{\arraystretch}{1.4}
\setlength{\tabcolsep}{15pt}
\begin{tabular}{|l| r r r|} 
\hline 
&  min. & default & max.\\  [0.5ex]  \hline
$\mathrm{M_{h}} [10^{12}\Ms]$ & $0$ & $1$ & $3$ \\ 
$\mathrm{M_{d}} [10^{10}\Ms]$ &  $4.2$ & $5.2$ & $6.2$ \\
$\mathrm{M_{b}} [10^{10}\Ms]$ & $0.5$ & $1$ & $3$ \\
$\mathrm{M_{n}} [10^{9}\Ms]$ & $0.5$ & $1$ & $3$ \\
$\mathrm{c_{NFW}}$ & 6 & 11 & 16 \\ 
 \hline
\end{tabular}
\end{center}
\tablefoot{Minimum, default, and maximum assumed parameter values of the gravitational potential explored here, comprising a Miyamoto-Nagai disk of mass $\mathrm{M_{d}}$, a bulge of mass $\mathrm{M_{b}}$, a nucleus of mass $\mathrm{M_{n}}$, and an NFW halo of mass $\mathrm{M_{h}}$ and concentration $\mathrm{c_{NFW}}$. The Miyamoto-Nagai disk scale lengths are fixed to $a_d = 3.0$~kpc and $b_d = 0.28$~kpc.}
\end{table}

\begin{table}
\caption{Age recovery with `next' errors}
\vspace{-3mm}
\label{tab:age-recovery}
\begin{center}
\renewcommand{\arraystretch}{1.4}
\setlength{\tabcolsep}{5pt}

\begin{tabular}{|l c| c c c c|}
\hline
\multicolumn{2}{|c|}{\multirow{2}{*}{\shortstack[l]{\bf Intrinsic\\ \bf properties}}} & $\mathrm{det}_\mathrm{min}$ & $\mathrm{tr}_\mathrm{min}$ & $\tau_\mathrm{det}$ & $\tau_\mathrm{tr}$ \\
\multicolumn{2}{|c|}{} & (pc) & (pc) & (Myr) & (Myr) \\
\hline
      & 10  &   $2.6^{+0.7}_{-0.6}$ & $6.3^{+1.8}_{-1.4}$ & $39.9^{+0.9}_{-0.9}$ & $39.6^{+1.2}_{-1.2}$ \\
$N_\star$ & 20   &  \boldmath$3.0^{+0.5}_{-0.5}$ & \boldmath$6.7^{+1.2}_{-1.2}$ & \boldmath$39.9^{+0.6}_{-0.6}$ & \boldmath$39.6^{+0.9}_{-0.8}$\\
      & 30     & $3.1^{+0.4}_{-0.4}$ & $6.9^{+1.0}_{-1.0}$ & $39.9^{+0.5}_{-0.5}$ & $39.5^{+0.7}_{-0.7}$\\
\hline
           & 0.5 pc & $2.9^{+0.5}_{-0.5}$ & $6.6^{+1.2}_{-1.2}$ & $39.9^{+0.6}_{-0.5}$ & $39.5^{+0.8}_{-0.8}$\\
$\sigma_x$ & 1 pc   & \boldmath$3.0^{+0.5}_{-0.5}$ & \boldmath$6.7^{+1.2}_{-1.2}$ & \boldmath$39.9^{+0.6}_{-0.6}$ & \boldmath$39.6^{+0.9}_{-0.8}$\\
           & 2 pc   & $3.2^{+0.5}_{-0.5}$ & $6.9^{+1.4}_{-1.2}$ & $39.9^{+0.6}_{-0.6}$ & $39.6^{+0.9}_{-0.8}$\\
\hline
           & $1\,\mathrm{km\,s^{-1}}$   & $2.7^{+0.5}_{-0.4}$ & $6.1^{+1.2}_{-1.1}$ & $39.9^{+0.9}_{-0.8}$ & $39.2^{+1.1}_{-1.2}$\\
$\sigma_v$ & $1.5\,\mathrm{km\,s^{-1}}$ & \boldmath$3.0^{+0.5}_{-0.5}$ & \boldmath$6.7^{+1.2}_{-1.2}$ & \boldmath$39.9^{+0.6}_{-0.6}$ & \boldmath$39.6^{+0.9}_{-0.8}$\\
           & $2\,\mathrm{km\,s^{-1}}$   & $3.1^{+0.5}_{-0.5}$ & $7.1^{+1.3}_{-1.3}$ & $39.9^{+0.4}_{-0.5}$ & $39.7^{+0.7}_{-0.7}$\\
\hline
      & 20 Myr & $1.7^{+0.3}_{-0.3}$ & $3.7^{+0.5}_{-0.6}$ & $19.9^{+0.4}_{-0.4}$ & $19.8^{+0.4}_{-0.4}$\\
Age   & 40 Myr & \boldmath$3.0^{+0.5}_{-0.5}$ & \boldmath$6.7^{+1.2}_{-1.2}$ & \boldmath$39.9^{+0.6}_{-0.6}$ & \boldmath$39.6^{+0.9}_{-0.8}$\\
      & 60 Myr & $2.9^{+0.4}_{-0.4}$ & $11.8^{+2.2}_{-2.1}$ & $60.0^{+0.2}_{-0.2}$ & $58.7^{+1.4}_{-1.4}$\\
\hline
\multicolumn{6}{c}{}\\[-0.3ex] 
\hline
\multicolumn{2}{|c|}{\multirow{2}{*}{\shortstack[l]{\bf Effects of \\ \bf the potential}}}  & $\mathrm{det}_\mathrm{min}$ & $\mathrm{tr}_\mathrm{min}$ & $\tau_\mathrm{det}$ & $\tau_\mathrm{tr}$ \\
\multicolumn{2}{|c|}{} & (pc) & (pc) & (Myr) & (Myr) \\
\hline
         & $0.6 \times 10^{12}\Ms$ & $3.5^{+0.6}_{-0.5}$ & $7.6^{+1.4}_{-1.1}$ & $42.9^{+0.8}_{-0.7}$ & $41.2^{+1.1}_{-1.1}$\\
$M_h$    & $1.0 \times 10^{12}\Ms$  & \boldmath$3.0^{+0.5}_{-0.5}$ & \boldmath$6.7^{+1.2}_{-1.2}$ & \boldmath$39.9^{+0.6}_{-0.6}$ & \boldmath$39.6^{+0.9}_{-0.8}$\\
         & $1.5 \times 10^{12}\Ms$   & $2.9^{+0.5}_{-0.4}$ & $6.6^{+1.3}_{-1.0}$ & $38.5^{+0.5}_{-0.5}$ & $38.7^{+0.8}_{-0.7}$\\
\hline
                 & 7 & $4.1^{+0.6}_{-0.5}$ & $9.0^{+1.3}_{-1.2}$ & $44.6^{+1.0}_{-0.7}$ & $41.1^{+1.3}_{-1.3}$\\
$c_\mathrm{NFW}$ & 11   & \boldmath$3.0^{+0.5}_{-0.5}$ & \boldmath$6.7^{+1.2}_{-1.2}$ & \boldmath$39.9^{+0.6}_{-0.6}$ & \boldmath$39.6^{+0.9}_{-0.8}$\\
                 & 15   & $3.3^{+0.5}_{-0.5}$ & $9.0^{+1.5}_{-1.5}$ & $35.9^{+0.4}_{-0.4}$ & $37.4^{+0.8}_{-0.9}$\\
\hline
           & $4.2  \times 10^{10}\Ms$  & $4.7^{+0.6}_{-0.6}$ & $9.4^{+1.2}_{-1.3}$ & $41.4^{+1.8}_{-4.1}$ & $41.0^{+1.2}_{-1.4}$\\
$M_d$ & $5.2  \times 10^{10}\Ms$ & \boldmath$3.0^{+0.5}_{-0.5}$ & \boldmath$6.7^{+1.2}_{-1.2}$ & \boldmath$39.9^{+0.6}_{-0.6}$ & \boldmath$39.6^{+0.9}_{-0.8}$\\
  & $6.2  \times 10^{10}\Ms$  & $3.7^{+0.5}_{-0.5}$ & $8.2^{+1.3}_{-1.3}$ & $35.7^{+0.7}_{-1.4}$ & $37.8^{+0.9}_{-1.0}$\\
\hline

\end{tabular}
\end{center}
\tablefoot{
Median values of $\mathrm{det}_\mathrm{min}$, $\mathrm{tr}_\mathrm{min}$, $\tau_\mathrm{det}$, and $\tau_\mathrm{tr}$, and their respective $\pm 1 \sigma$ equivalent uncertainties, assuming `next' observational errors.  Bold values for a given parameter are equal across both tables, because they share the same default combination of intrinsic parameters and potential parameters. The top table shows results for associations of different intrinsic parameters, evolved backwards, assuming that the true potential is known and equal to the default potential. The uncertainty or bias of the inferred values is purely due to observational errors.
The bottom table shows results for associations of the default intrinsic parameters, but evolved backwards in different potentials. The uncertainty or  bias on the inferred values can result from a combination of observational errors and uncertainty about the potential.}
\end{table}

\section{Varying the potential} \label{sec:potential}
In the previous section, we assumed that the gravitational potential was known, i.e. the backwards integration was always performed in the same potential in which the stellar associations had been integrated forwards in time. We then varied the potential for the backwards integration and used two strategies to constrain the true potential: one based on the association's minimum size (Sect.~\ref{sec:min-size}) and one based on the association's age (Sect.~\ref{sec:age-inference}). 

We assumed that the true potential had the default values given in Table~\ref{tab:potential} and varied each parameter in turn over the range given in the same table. We found the effects of the masses of the bulge and nucleus on associations in the solar neighbourhood to be so small that we adopted their default values for all results shown.

Since we can currently apply the traceback method only in the solar neighbourhood, as discussed in Section~\ref{sec:errors}, our inference of the potential or its components is effectively a local measurement near the solar circle, $R_0$. Inference of the Galactic potential is therefore contingent on its parametrisation.

\subsection{Constraining the potential from the association's minimum size}~\label{sec:min-size}

\begin{figure*}
\centering

    \includegraphics[clip,trim=0 1.4cm 0 0.6cm,width=2.31in]{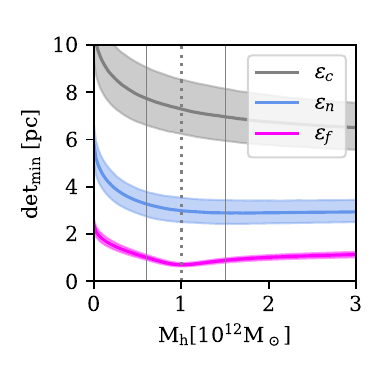}
    \includegraphics[clip,trim=0 1.4cm 0 0.6cm,width=2.31in]{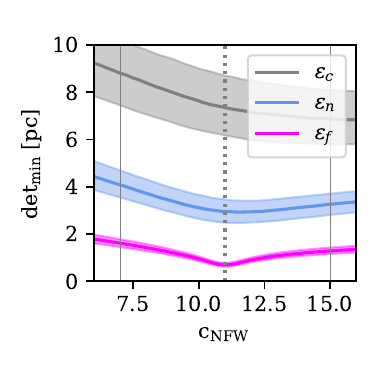}
    \includegraphics[clip,trim=0 1.4cm 0 0.6cm,width=2.31in]{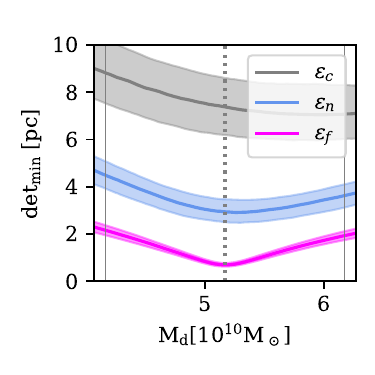} \\   
    \includegraphics[clip,trim=0 0.5cm 0 0.6cm,width=2.31in]{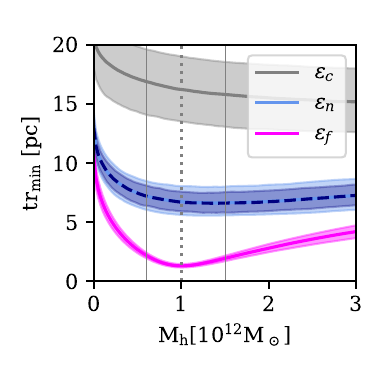}
    \includegraphics[clip,trim=0 0.5cm 0 0.6cm,width=2.31in]{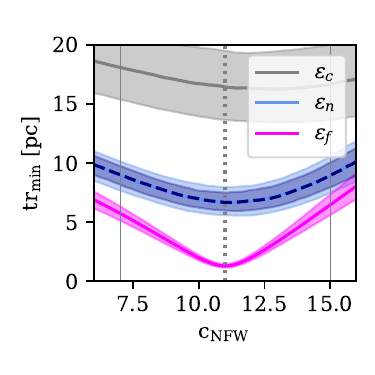}
    \includegraphics[clip,trim=0 0.5cm 0 0.6cm,width=2.31in]{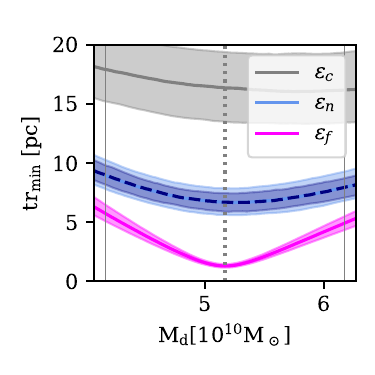} \\

    \caption{Sensitivity of the association's minimum size to components of the Galactic potential. We show the median and $1 \sigma$ dispersion of the minimum value of the determinant (top row) and the trace (bottom row). A different potential component is analysed in each column, namely the halo mass (left), halo concentration (centre), and disk mass (right). Vertical dotted lines indicate the values of the true potential. As shown by the dashed dark blue line in the bottom row, averaging over the results of multiple associations can further reduce the uncertainty.}
   \label{fig:sensitivity-size}
\end{figure*}

\begin{figure*}
\centering
   \includegraphics[clip,trim=1.cm 0.5cm 0.2cm 0.6cm,width=5.95cm]{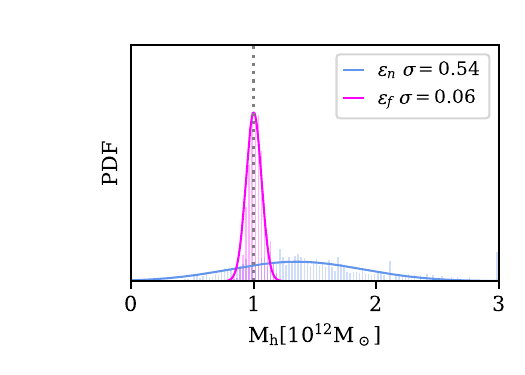}
   \includegraphics[clip,trim=1.cm 0.5cm 0.2cm 0.6cm,width=5.95cm]{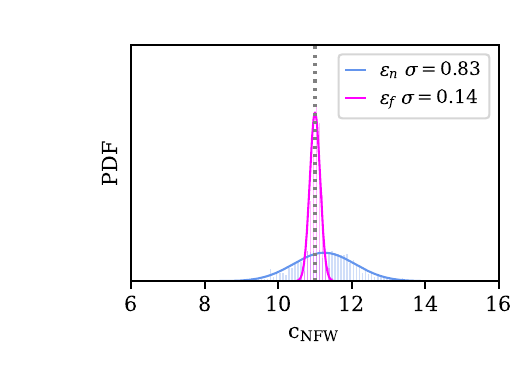}
   \includegraphics[clip,trim=1.cm 0.5cm 0.2cm 0.6cm,width=5.95cm]{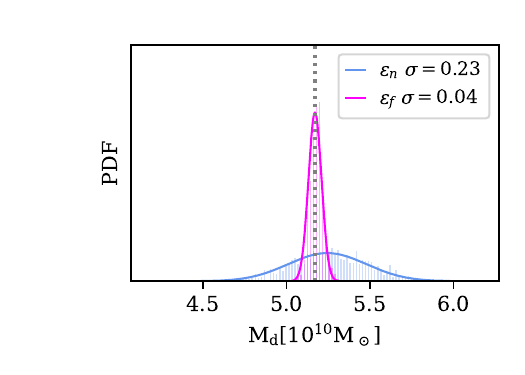} \\
  
   \caption{Posterior distributions for potential components inferred from single associations, assuming uniform priors over the range indicated. Obtained from the smallest minimum trace using realisations of the default association with `next', $\epsilon_n$, or `future', $\epsilon_f$, errors. Dotted vertical lines indicate the values of the true potential.}
   \label{fig:results-posterior-single-parameter}
\end{figure*}

Figure~\ref{fig:sensitivity-size} shows the minimum size of the association, quantified by $\mathrm{det}_\mathrm{min}$ and $\mathrm{tr}_\mathrm{min}$, as the potential is varied in the backwards integration. We performed this analysis for the three most important components of the potential, namely the halo mass, $\mathrm{M_h}$, halo concentration, $c_\mathrm{NFW}$, and disk mass $M_\mathrm{d}$. In each case, we assumed the default association's intrinsic parameters, described in Table~\ref{tab:intrinsic}, and we show results for the three data scenarios `current' ($\epsilon_c$), `next' ($\epsilon_n$), and `future' ($\epsilon_f$) errors, described in Table~\ref{tab:errors}. We integrated the association forward in the default potential and separately varied each parameter in the backward integration. The dotted vertical lines show the true values of the parameters, i.e. those assumed in the forward integration. 

As noted previously, it is not the value itself, but the local curvature around the minimum relative to the variance, that allows inference of the component. In general, the minimum value of the trace, $\mathrm{tr}_\mathrm{min}$, shows a greater sensitivity to the potential than the minimum value of the determinant, $\mathrm{det}_\mathrm{min}$. Therefore, the minimum value of the trace allows us to constrain the potential parameters more precisely than the determinant.

It is also evident that, with `current' errors, the dependences of either the trace or determinant on the potential are too weak to provide a meaningful inference of any of the components of the potential. Additionally, the smallest size is not always inferred in the true potential, possibly introducing biases in the inferred potential parameters. However, with `next' errors, the average minima of the trace become sensitive to all three components of the assumed potential, and the minimum occurs in the true potential, albeit with a rather shallow dependence on the halo mass.

Figure~\ref{fig:results-posterior-single-parameter} shows the posterior distributions for parameters of the Galactic potential inferred from the smallest minimum trace, obtained after uniformly sampling individual parameters of the potential in the backwards integration over the intervals shown, while keeping the other parameters fixed. Histograms show the raw data from 1000 samples, lines show Gaussian fits to the distribution. The standard deviation, $\sigma$, of the Gaussian fit is also shown.

It can be seen that `current' errors are generally not sufficient to infer the true values of the halo mass, and as suggested by Figure~\ref{fig:sensitivity-size}, with `current' errors, the maximum sampled halo mass ($3\times 10^{12}\Ms$) often results in the smallest trace. When the halo mass is fixed, however, the concentration and the disk mass may be meaningfully constrained, even assuming `current' errors, at least under the idealised conditions we consider here.

With `next' errors, the posterior distributions peak at or near the values of the true potential for all three components. Assuming that the other parameters of the density distribution are fixed, meaningful constraints can be placed on each parameter individually. For a single association, the Gaussian fits give approximate uncertainties of $\sigma(M_h) = 5.4 \times 10^{11}~\Ms$, $\sigma(c_\mathrm{NFW}) = 0.83$, and $\sigma(M_d) = 2.3 \times 10^{9}~\Ms$ with `next' errors; these improve to $6 \times 10^{10}~\Ms$, $0.14$, and $4 \times 10^{8}~\Ms$, respectively, with `future' errors. As shown for the example of three associations with `next' errors, combining the results of multiple associations, all assumed to evolve in the same potential, can further reduce the uncertainty. However, in part due to the degeneracies described below, these values should be interpreted as idealised one-parameter constraints.

For comparison, \citet{Cautun-2020} obtained $M_{200} = 0.97_{-0.19}^{+0.24} \times 10^{12}~\Ms$ and $c =9.4^{+1.9}_{-2.6}$ from the Milky Way's rotation curve, while \citet{Callingham-2019} inferred $M_{200} = 1.17^{+0.21}_{-0.15} \times 10^{12}~\Ms$ and $c=10.9^{+2.6}_{-2.0}$ from satellite dynamics. However, as emphasised in \citet{Hunt-2025}, rotation curves and larger-scale enclosed-mass measurements are not fully consistent. Expanding associations should not be viewed as a replacement for other tracers, but a complementary tool with different systematics.

\begin{figure*}
\centering
   \includegraphics[width=6.1cm,trim={0cm 1.3cm  2.2cm 0.8cm},clip]{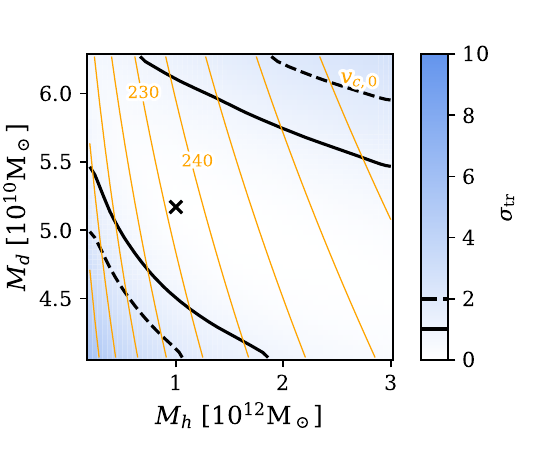}
   \includegraphics[width=5.0cm,trim={1.25cm 1.3cm  2.2cm 0.8cm},clip]{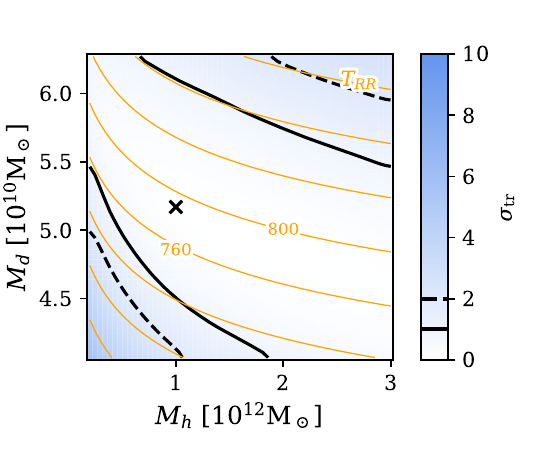}
   \includegraphics[width=6.93cm,trim={1.25cm 1.3cm 0cm 0.8cm},clip]{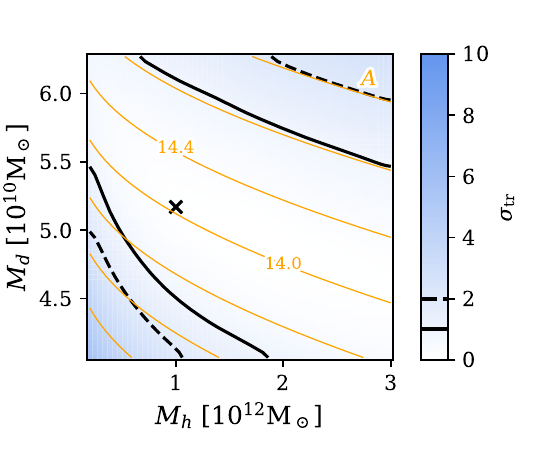}

   \includegraphics[width=6.1cm,trim={0cm 0cm  2.2cm 0.8cm},clip]{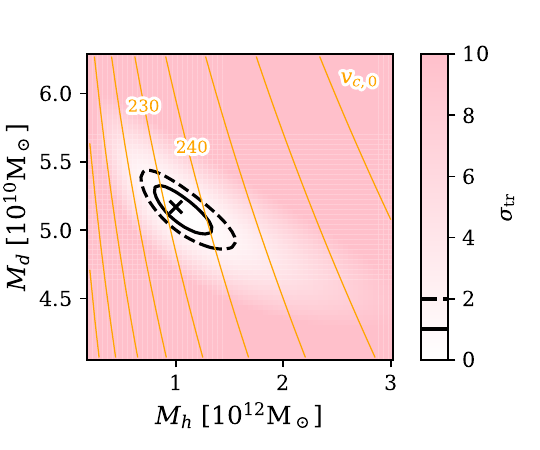}
   \includegraphics[width=5cm,trim={1.25cm 0cm  2.2cm 0.8cm},clip]{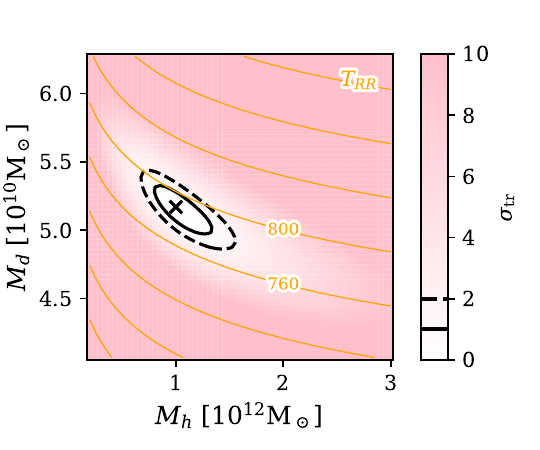}
   \includegraphics[width=6.93cm,trim={1.25cm 0cm 0cm 0.8cm},clip]{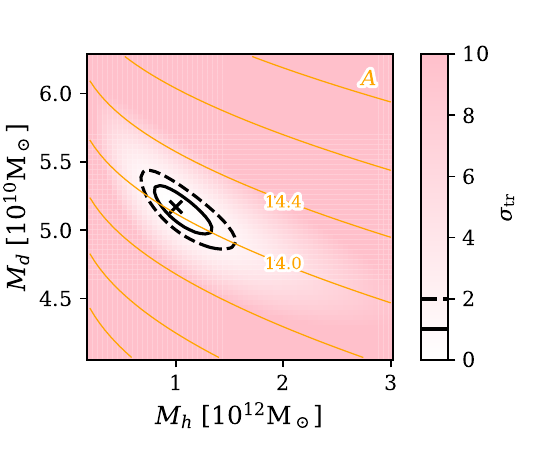}
   
   \includegraphics[width=6.1cm,trim={0cm 1.3cm  2.2cm 0.8cm},clip]{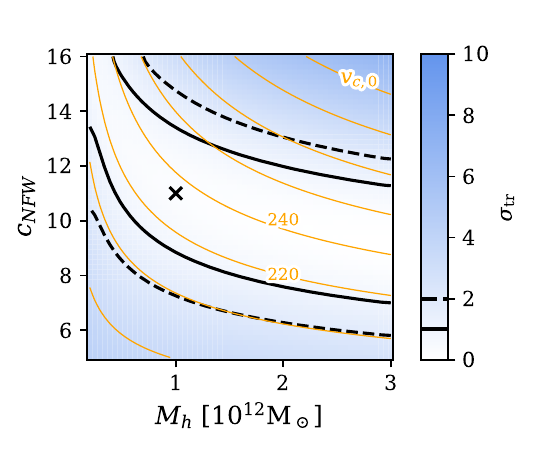}
   \includegraphics[width=5cm,trim={1.25cm 1.3cm  2.2cm 0.8cm},clip]{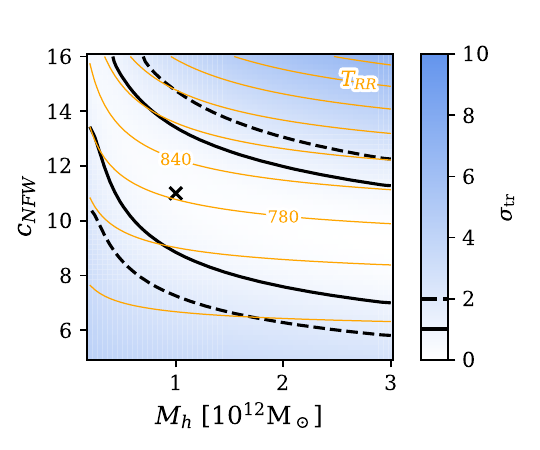}
   \includegraphics[width=6.93cm,trim={1.25cm 1.3cm 0cm 0.8cm},clip]{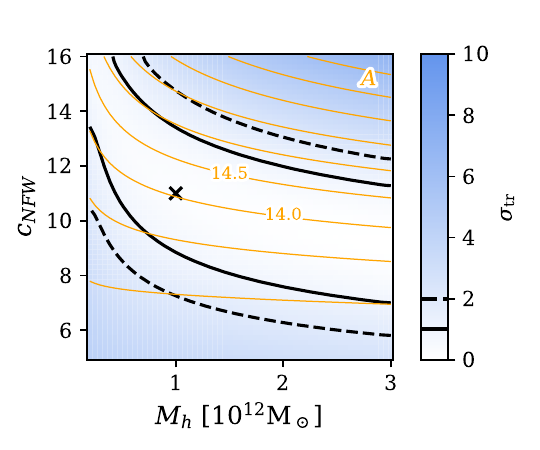}

   \includegraphics[width=6.1cm,trim={0cm 0cm  2.2cm 0.8cm},clip]{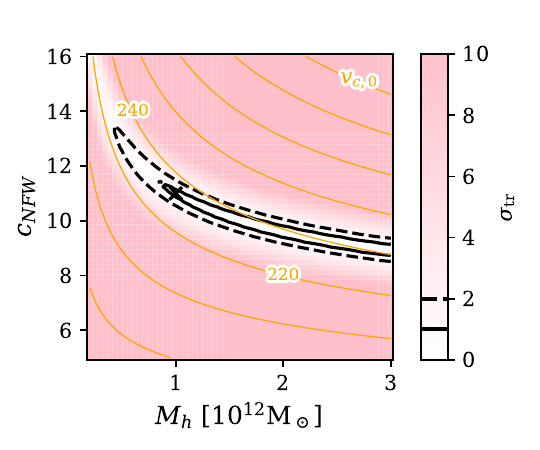}
   \includegraphics[width=5cm,trim={1.25cm 0cm  2.2cm 0.8cm},clip]{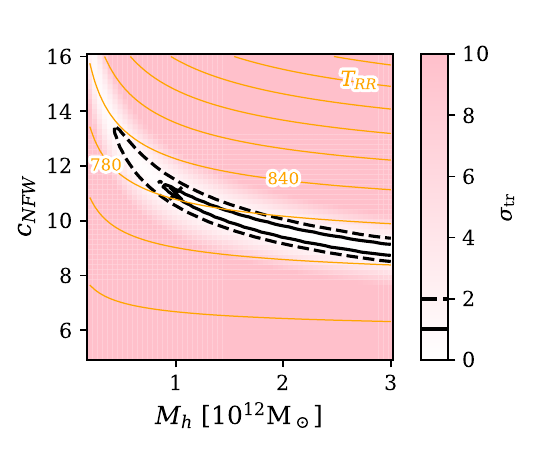}
   \includegraphics[width=6.93cm,trim={1.25cm 0cm 0cm 0.8cm},clip]{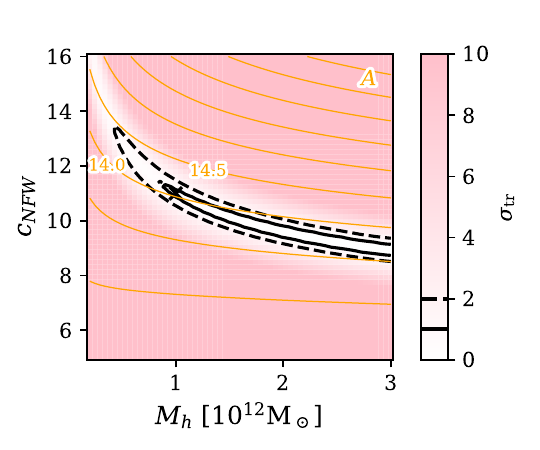}

   \caption{Degeneracy between  pairs of potential variables, disk mass and halo mass (top two rows), and concentration and halo mass (bottom two rows) for inferring the Galactic potential from the minimum of the trace using a single expanding association with `next' errors (blue) or `future' errors (pink).
   On each panel, the cross marks the parameters of the true potential, solid and dashed lines mark confusion regions of 1$\sigma$ and 2$\sigma$, respectively. Orange contours show lines of constant $v_{c,0}$, constant $T_{RR}$, and constant Oort constant $A$. At fixed $R_0$, the $v_{c,0}$ contours are equivalent to contours of constant local radial acceleration, $v_{c,0}^2/R_0$. Orange numbers show the value of the parameter for the two lines that enclose the minimum.
   }
   \label{fig:degeneracy}
   \vspace{0.4cm}   
\end{figure*}

\subsection{Degeneracy between potential parameters}

Different combinations of parameters can result in very similar tidal fields and accelerations near $R_0$, introducing degeneracy into the reconstruction.
In Figure~\ref{fig:degeneracy}, we show the degeneracies between pairs of variables, halo mass and disk mass, and halo mass and concentration, for the inference of the potential using the minimum of the trace, and based on a single association with either `next' or `future' errors. On each panel, we show the median of the minimum of the trace, for an association forward-integrated in the default potential, and backwards integrated in a potential with the corresponding combination of parameters, with all remaining parameters held constant. The colour scales are normalised by the smallest average value of the minimum trace. The parameter choices of the assumed true potential are indicated by the black cross, contours show confusion regions, for which the median values of the minimum trace are consistent with the absolute minimum at $1 \sigma$ or $2 \sigma$.
The shapes of the confusion regions are as expected: a higher concentration or a higher disk mass in combinations with a lower halo mass can yield similar values of the minimum trace.

On each panel, we also overplot representative derived local quantities, namely the circular velocity at the solar radius, $v_{c,0}$, the radial component of the local tidal tensor, $T_{RR}$, and the Oort constant, $A$. Since $R_0$ is fixed, contours of constant $v_{c,0}$ are equivalent to contours of constant local radial acceleration, $v_{c,0}^2/R_0$. We show the values of these quantities for the two lines that enclose the value of the minimum trace; the remaining lines use the same spacing on each panel.

In the $M_h - M_d$ plane, the shape of the confusion region is roughly aligned with $A$ and $T_{RR}$, but runs perpendicular to the $v_{c,0}$ contours. While the circular velocity depends strongly on the halo mass, the local tidal field is also significantly influenced by the disk.
In the $M_h - c$ plane, where both parameters control the spherical halo, the shape of the confusion region is most closely aligned with regions of constant $v_{c,0}$, i.e. constant mass interior to $R_0$.

Even with `next' errors, a single association cannot be used to place meaningful constraints on all components of the potential simultaneously. On the other hand, if some of the parameters can be constrained independently, the remaining ones can be inferred.

\subsection{Effect of the potential on the traceback age} \label{sec:age-effect}

\begin{figure*}
\centering

    \includegraphics[clip,trim=0 1.4cm 0 0.6cm,width=2.31in]{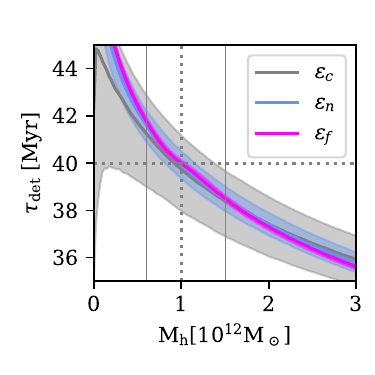}
    \includegraphics[clip,trim=0 1.4cm 0 0.6cm,width=2.31in]{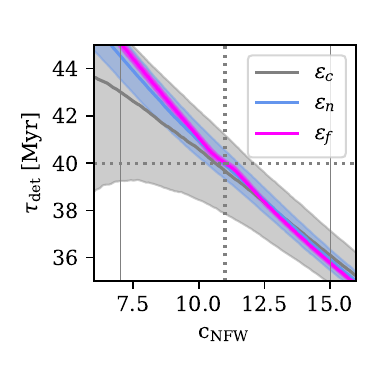}
    \includegraphics[clip,trim=0 1.4cm 0 0.6cm,width=2.31in]{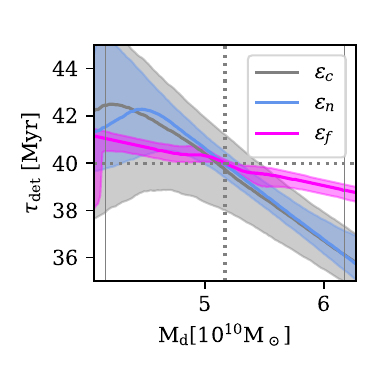} \\    
    \includegraphics[clip,trim=0 0.5cm 0 0.6cm,width=2.31in]{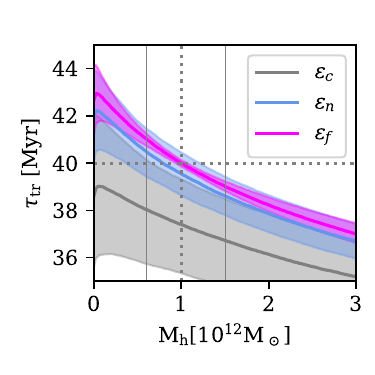}
    \includegraphics[clip,trim=0 0.5cm 0 0.6cm,width=2.31in]{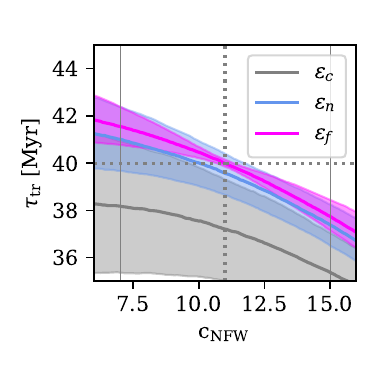}
    \includegraphics[clip,trim=0 0.5cm 0 0.6cm,width=2.31in]{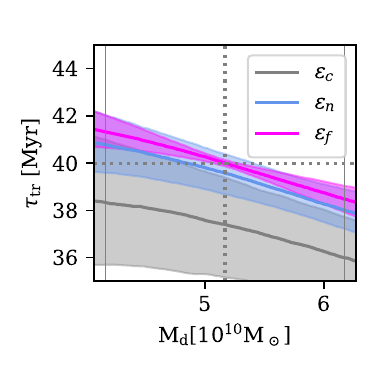} \\

   \caption{Sensitivity of the traceback age to individual components of the potential. On each panel, we show the median and $\pm 1 \sigma$ dispersion of the inferred age using either the determinant (top row) or trace (bottom row). All results are for the default association, assuming either `current' ($\epsilon_c$; grey), `next' ($\epsilon_n$; blue), or `future' ($\epsilon_f$; pink) errors. Dotted horizontal and vertical lines show the true age and parameter value in the true potential, respectively. Solid horizontal lines indicate the parameter values detailed in Table~\ref{tab:age-recovery}. In the true potential, for a given set of observational errors, the determinant provides a more accurate and less biased traceback age. However, the age inferred from the determinant is more sensitive to the potential than the age inferred from the trace.}
   \label{fig:sensitivity-age}
\end{figure*}

\begin{figure*}
\centering
    \includegraphics[clip,trim=1.cm 1.5cm 0.2cm 0.6cm,width=5.95cm]{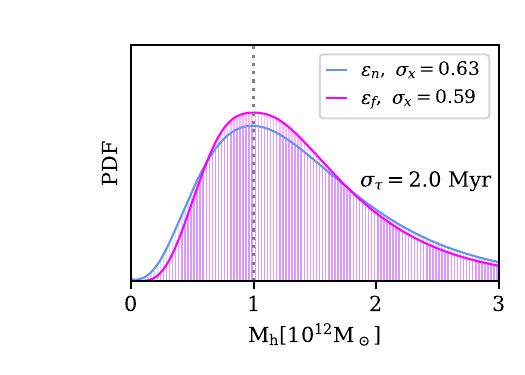}
    \includegraphics[clip,trim=1.cm 1.5cm 0.2cm 0.6cm,width=5.95cm]{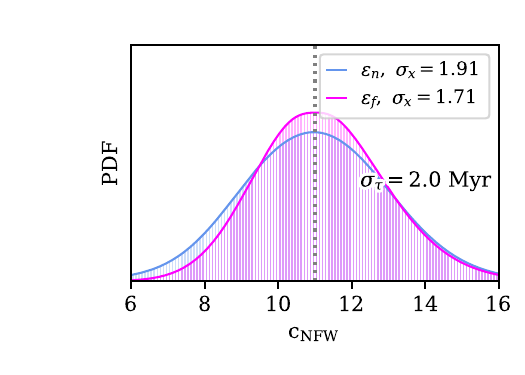}
    \includegraphics[clip,trim=1.cm 1.5cm 0.2cm 0.6cm,width=5.95cm]{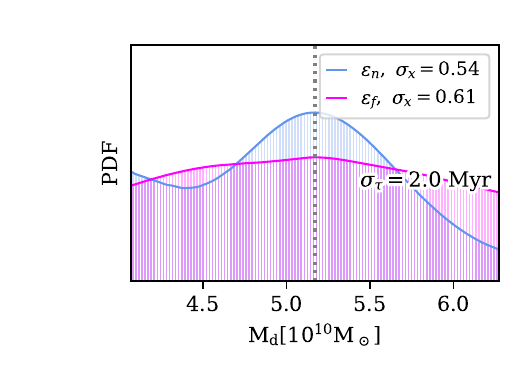}
    \includegraphics[clip,trim=1.cm 1.5cm 0.2cm 0.6cm,width=5.95cm]{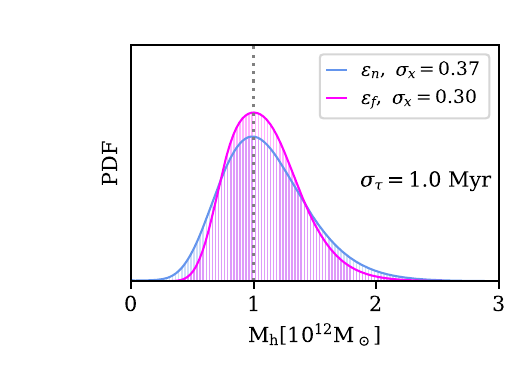}
    \includegraphics[clip,trim=1.cm 1.5cm 0.2cm 0.6cm,width=5.95cm]{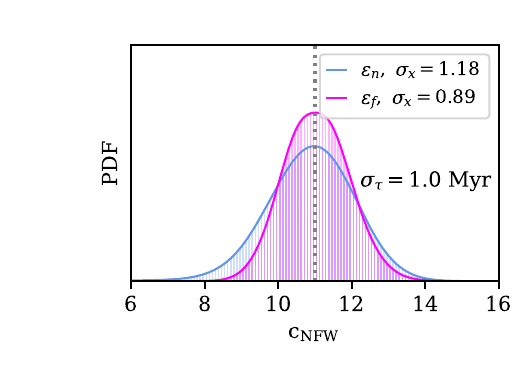}
    \includegraphics[clip,trim=1.cm 1.5cm 0.2cm 0.6cm,width=5.95cm]{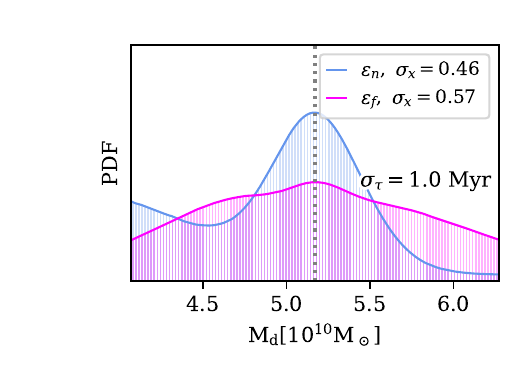}
    \includegraphics[clip,trim=1.cm 0.5cm 0.2cm 0.6cm,width=5.95cm]{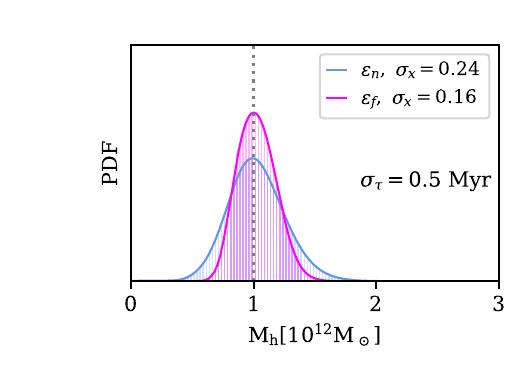}
    \includegraphics[clip,trim=1.cm 0.5cm 0.2cm 0.6cm,width=5.95cm]{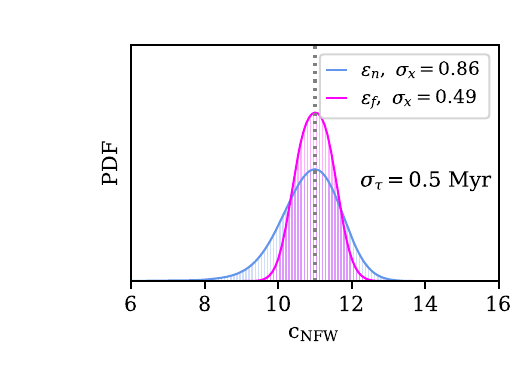}
    \includegraphics[clip,trim=1.cm 0.5cm 0.2cm 0.6cm,width=5.95cm]{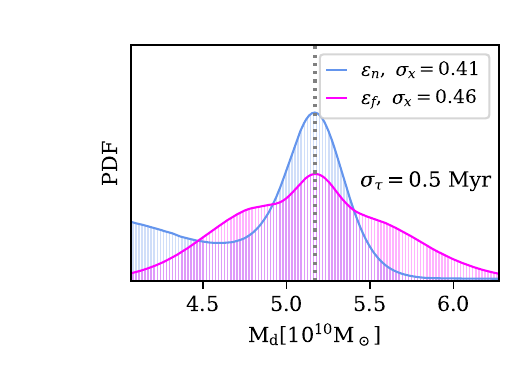}
    
   \caption{Posterior distributions for potential components inferred from single associations, assuming uniform priors over the range indicated, analogous to Figure~\ref{fig:results-posterior-single-parameter}. Here, we use $\tau_\mathrm{det}$ and assume that the independent age estimate is $\mu_\tau = 40$~Myr, with a precision of $\sigma_\tau = 2.0$~Myr (top), $\sigma_\tau = 1.0$~Myr (middle), or $\sigma_\tau = 0.5$~Myr (bottom row). A precise independent age determination, in combination with `next' or `future' errors, can provide meaningful constraints on the parameters of the potential.}
   \label{fig:results-posterior-single-parameter-age}
\end{figure*}

Figure~\ref{fig:sensitivity-age} shows the sensitivity of the traceback ages, $\tau_\mathrm{det}$ and $\tau_\mathrm{tr}$, to the three components of the potential investigated in this work, namely the halo mass, $\mathrm{M_h}$, halo concentration, $c_\mathrm{NFW}$ and disk mass $M_\mathrm{d}$. In each case, we assumed the default association's intrinsic parameters, described in Table~\ref{tab:intrinsic}, and we show results for the three data scenarios `current' ($\epsilon_c$), `next' ($\epsilon_n$), and `future' ($\epsilon_f$) errors, described in Table~\ref{tab:errors}. We integrated the association forward in the default potential and separately varied each parameter in the backward integration.

Dotted vertical lines show the true values of the parameters, i.e. those assumed in the forward integration. Dotted horizontal lines show the true age of the association, 40 Myr. The top row shows the dynamical traceback age of the association inferred using the determinant, the bottom row shows results using the trace.

Reassuringly, when the backwards integration is performed in the true potential, both metrics recover traceback ages close to the true age in all cases. Using `current' errors, in the true potential, $\tau_{\mathrm det}$ is essentially unbiased and has a smaller uncertainty, while $\tau_{\mathrm tr}$ shows a bias of $\sim 2$ Myr and a larger uncertainty.

As expected, the bias and the uncertainty in both $\tau_\mathrm{det}$ and $\tau_\mathrm{tr}$ decrease when adopting smaller errors. However, we note that $\tau_\mathrm{det}$ is also significantly more sensitive than $\tau_\mathrm{tr}$. For associations of this age, the systematic uncertainty of the traceback age caused by uncertainties in the potential can be as large as the statistical uncertainty due to `current' measurement errors, and it can become the dominant source of uncertainty when the reduced `next' measurement errors are assumed. This effect is less severe for younger associations and more severe for older associations.

\subsection{Constraining the potential from the traceback age} \label{sec:age-inference}
The fact that the inferred traceback age depends on the potential can also be used to infer the potential, if an independent age measurement exists: the correct potential is the one in which the traceback age best matches the independently measured age, $\tau_\mathrm{ind}$. In practice, any independent age measurement has an associated uncertainty. If we treat $\tau_\mathrm{ind}$ as a normal distribution with mean, $\mu_\tau$, and standard deviation, $\sigma_\tau$, the likelihood that a given traceback age measurement, $\tau_\mathrm{det}$, resulting from an assumed potential, $\Phi$, is consistent with the observed $\tau_\mathrm{ind}$, is:

\[
\mathcal{L}(\tau_{\mathrm ind} \mid \Phi)
\propto
\exp\!\left[
-\frac{\bigl( \tau_\mathrm{det}(\Phi) -\mu_\tau\bigr)^2}{2\sigma_\tau^2}
\right].
\]

Assuming a uniform prior on a parameter (such as the mass or concentration) of the potential, $x$, the posterior for $x$ is obtained by averaging the likelihood over the Monte Carlo realisations of the association,
\begin{equation}
p(x \mid \tau_{\mathrm ind}) \propto \frac{1}{N_{\mathrm MC}} \sum_{i=1}^{N_{\mathrm MC}}
\exp\!\left[-\frac{\left(\tau_{\mathrm{det},i}(x)-\mu_\tau\right)^2}{2\sigma_\tau^2}\right].
\end{equation}
In other words, the potential most consistent with the independent age measurement is the one that yields the most consistent traceback age, taking into account the shape of $\tau_\mathrm{det}(x)$ and the uncertainty of the independent age measurement.

In Figure~\ref{fig:results-posterior-single-parameter-age}, we show the posterior distributions, $p(x)$, for components of the potential, assuming an association with the default parameters, including a true age of 40~Myr, observed with different error budgets, and with uncertainties in the independent age measurement of $\sigma_\tau = 2$~Myr, $\sigma_\tau = 1$~Myr, or $\sigma_\tau = 0.5$~Myr. In these idealised scenarios, and assuming `next' errors, the sensitivity of the traceback age to the potential can constrain components of the potential, particularly the halo mass and concentration. By comparison, as was shown in Figure~\ref{fig:sensitivity-age}, at least over the interval considered, the mass of the disk has a smaller and less monotonic relationship to the traceback age, which translates to a reduced power to constrain the parameter using an independent age measurement.

\section{Discussion}

\subsection{Perspectives for dynamical traceback ages}
The idea of rewinding stellar motions, i.e. inferring past positions from present-day velocities, is very old, but its practical application has long been limited by the lack of precise 6D phase-space information for individual stars. Hipparcos provided the first opportunity to reconstruct stellar trajectories in nearby young associations, enabling estimates of their dynamical ages and birth sites in some favourable cases \citep{Song-2003, Fernandez-2008}. The arrival of Gaia, with its outstanding improvement in astrometric precision for over a billion stars, has led to a resurgence of orbital traceback studies \citep{Miret-Roig-2020, Miret-Roig+2022, Kerr+2022, Galli-2023, Couture+2023}.

Our results show that dynamical traceback ages based on Gaia DR3 astrometry and precise radial velocities can already achieve a high precision comparable to that obtained from isochrone fitting. This opens the possibility of using traceback ages as an independent tool to constrain and calibrate pre-main-sequence evolutionary models. Moreover, comparing these two age-dating techniques provides valuable information about the early dynamical evolution of stellar systems, including the role of molecular cloud dispersal in driving the expansion of young clusters \citep{Miret-Roig-2024}.

While astrometric precision is expected to continue to improve with forthcoming Gaia data releases and future missions such as GaiaNIR, it is essential that radial-velocity precision advances in parallel. Currently, instruments like HARPS and ESPRESSO can already provide m/s radial velocity precision, but this is only possible for very few stars. Modern spectroscopic surveys, including SDSS-V \citep{Kollmeier-2019}, 4MOST \citep{deJong+2019}, and WEAVE \citep{Dalton+2014}, provide large, homogeneous radial velocity catalogues; however, achieving precisions at the level of $\lesssim 0.2\,\mathrm{km\,s^{-1}}$ remains challenging. Future facilities, such as the Wide-field Spectroscopic Telescope (WST) \citep{Mainieri+2024} and Maunakea Spectroscopic Explorer (MSE) \citep{MSE+2019}, are expected to play a key role in reaching this regime. In parallel, it is critical to characterise and correct systematic offsets between different surveys \citep{Tsantaki+2022} and those arising from intrinsic stellar properties \citep{Couture+2023, Gagne+2026} to fully exploit these improvements and achieve the desired radial-velocity precision.

\subsection{Inference of the potential from young stellar associations}
Our results show that young stellar associations that have expanded under the influence of the local gravitational potential can, in principle, be used to constrain the Galactic potential, or equivalently, derived local quantities such as $v_{c,0}^2/R_0$ and components of the local tidal tensor. While `current' observational errors are not sufficient, the improved astrometry expected from Gaia DR4, combined with higher-precision RV measurements, can provide complementary constraints on the dark matter density near the Sun. These can either offer independent validation of existing measurements, such as those obtained from the rotation curve  \citep[e.g.][]{Cautun-2020, Karukes-2020, Ablimit-2020, Shen-2022}, stellar streams \citep[e.g.][]{PW-2014, Gibbons-2014, Palau-2025}, orbits of globular clusters \citep[e.g.][]{Watkins-2019, Posti-2019} or satellite galaxies \citep[e.g.][]{Callingham-2019, Fritz-2020, Li-2020, Rodriguez-Wimberly-2022}, abundance matching \citep[e.g.][]{Guo-2010}, or the Timing Argument \citep[e.g.][]{Benisty-2022, Sawala-2023b}, or be used in combination with other methods to provide greater precision, particularly for parameters such as the halo concentration that strongly affect the potential in the solar neighbourhood.

Our results also show that uncertainties about the assumed Galactic potential for the orbital integration can affect the accuracy of dynamical traceback ages (see Table~\ref{tab:age-recovery}). While observational errors are the currently dominant source of uncertainty for the traceback age, uncertainty about the potential can become important when the observational errors decrease. As a corollary, when the true age of the association is known independently, one can also use this dependence of the dynamical traceback age to infer the parameters of the potential.

Compared to the conceptual analogue of inference from globular cluster streams, expanding associations are much younger, with typical lifetimes of a few tens of millions of years, rather than the billions of years of globular cluster streams \citep{Bose-2018}. While the evolution of a globular cluster stream is affected by (and can potentially be used to infer) the history of the evolving Galactic potential, young stellar associations offer a snapshot of its close-to-present state, much more localised in both time and position.

We have focused here on individual associations, but combining multiple associations and incorporating independent age information can tighten constraints and reduce degeneracies between potential components, in close analogy to multi-stream inference \citep[e.g.][]{Bonaca-2025,Palau-2025}. Alternatively, one could consider cluster families, i.e. groups of associations that share a common origin \citep{Swiggum-2024}. In this instance, instead of individual stellar orbits whose reconstruction requires excellent radial-velocity precision for every star, radial velocities can be averaged over all stars in an association. 

While foreseeable observational precision limits the analysis to associations in the solar neighbourhood, further improved parallaxes could extend the method to more distant ones. `current' errors give a distance error of $\sim 2$~pc at a distance of $\sim 200$~pc, while the `next' and `future' error budgets indicated in Table~\ref{tab:errors} give comparable precision at distances of $\sim 500$~pc and $\sim1$~kpc, respectively. Using the catalogue of \cite{Hunt-2023}, this corresponds to increasing the number of known young stellar associations from 150 to about 400 and 900, respectively. The inclusion of more distant associations would also reduce the reliance on parametrisations and help break existing degeneracies.

\subsection{Caveats}
The main caveat we have identified is that the current observational precision is not yet sufficient to provide meaningful constraints on the potential from young expanding associations. At present, and even more so with imminent improvements to astrometry, radial velocities are the limiting factor, making precise homogeneous radial velocity surveys crucial for the reconstruction of stellar orbits.

In addition, throughout this work, we make a number of simplifying assumptions. While many of them are reasonably well motivated, others may need to be revisited once our method is applied to real data.

{In this proof-of-concept}, we assume that the potential is static and axisymmetric. This may be justified due to the relatively young age of the associations compared to characteristic time scales for the evolution of the structures such as bars and spiral patterns, which typically vary on time scales of $\gtrsim 10^8$ yr \citep[e.g.][]{Sellwood-2014, Sellwood-2022}. Non-equilibrium structures and motions within the disk \citep[e.g.][]{Antoja-2023,Grosbol-2018} may need to be considered, particularly for older associations.

We also assume that the halo is spherical, and that the contribution from the disk is due to a smooth, single Miyamoto-Nagai disk, whose mass we assume to be its only free parameter. Especially the latter is certainly not accurate \citep[e.g.][]{Antoja-2018}. In its present form, our method should be interpreted as constraining combinations of the local acceleration field, including the local radial acceleration $v_{c,0}^2/R_0$, and its spatial gradients, i.e. the local tidal tensor. Constraints on the potential, or on the components of the underlying density model, are therefore dependent on the assumed parametrisation. The method could easily be extended to introduce more complex and variable potentials, albeit at the price of further degeneracies.

In considering idealised associations, we ignore the effect of binary stars, anisotropic phase-space distributions, rotation, completeness, and contamination. In practice, precise radial velocities can help to identify spectroscopic binaries and other kinematic contaminants \citep[e.g.][]{Gagne-2018, Miret-Roig-2020, Olivares-2025}. We have also explored how independent knowledge of the association's age can help constrain the Galactic potential. However, some discrepancies have been measured between isochronal and traceback ages, which should be considered \citep{Miret-Roig-2024}. On the other hand, it is possible that more sophisticated metrics, which take into account the full phase-space distribution rather than the trace of the covariance matrix, are better able to discriminate between different potentials. We leave this to future work.

\section{Conclusion}
We have explored how young stellar associations can, in principle, constrain the Galactic gravitational potential by integrating the orbits of their members back to the compact birth configuration. Both the minimum size of an association and its dynamical traceback age depend on the assumed potential. We have shown how the minimum size and, when combined with independent age measurements, the traceback age, can thus be used to  constrain the potential in which the association has evolved.

Our synthetic experiments indicate that the method has only limited constraining power at current observational precision, but becomes substantially more informative with upcoming astrometry and radial velocities. To take full advantage of the astrometric precision expected from future Gaia data releases, it is crucial to improve radial-velocity precision by about an order of magnitude, to the level of $\sim 0.2\,\mathrm{km\,s^{-1}}$. In the near term, the method benefits most from improved per-star precision rather than from larger sample sizes. In the long term, precise phase-space information for stars at larger distances could allow us to constrain the Galactic potential beyond the solar neighbourhood. This could strengthen the case for a successor to Gaia \citep{Hobbs-2016, Hobbs-2021}. 

Although significant degeneracies between potential parameters remain, expanding stellar associations offer a complementary way to infer the Galactic gravitational potential, particularly within the solar circle. At the same time, our results show that the potential is not merely the target of the inference, but can itself affect the inferred age: as observational uncertainties decrease, uncertainty in the assumed potential becomes an important, and in some cases limiting, source of uncertainty in dynamical traceback ages. This dual role motivates a joint treatment of the potential and of traceback ages.

\begin{acknowledgements}
We thank Adrian Price-Whelan and Kathryn Johnston for very helpful discussions, and we thank the reviewer for very helpful comments which have allowed us to improve the manuscript. T.S. thanks Flatiron Institute and the Simons Foundation for travel support. T.S. gratefully acknowledges support by the Research Council of Finland grants 354905 and 339127. N. M. R. acknowledges financial support from the Beatriu de Pinós postdoctoral fellowship (2023 BP 00215), awarded by the Agència de Gestió d'Ajuts Universitaris i de Recerca (AGAUR), Generalitat de Catalunya, and the Ramón y Cajal fellowship (RYC2024-051353-I), funded by MICIU/AEI/10.13039/501100011033 and by the European Social Fund Plus (ESF+). This work used facilities hosted by the CSC—IT Centre for Science, Finland. We also gratefully acknowledge the use of open-source software, including \texttt{Gala} \citep{Price-Whelan-2017}, \texttt{Matplotlib} \citep{matplotlib-paper}, \texttt{SciPy} \citep{SciPy} and \texttt{NumPy} \citep{numpy-paper}.
\end{acknowledgements}

\section*{Data availability}
Documented code to reproduce all results and figures presented in this paper is provided at: \\ \url{https://www.github.com/TillSawala/accelerometers}.

\bibliographystyle{aa} 
\bibliography{paper.bib} 

\end{document}